\documentclass[preprint2]{aastex}

\newcommand\KAB{K_{\rm AB}}
\newcommand\KVega{K_{\rm Vega}}
\newcommand\Btwosigma{B_{\rm 2\sigma\, lim}}
\newcommand\ztwosigma{z'_{\rm 2\sigma\, lim}}
\newcommand\xisBzK{\xi_{\rm sBzK}}
\newcommand\xiDM{\xi_{\rm DM}}
\newcommand\beff{b_{\rm eff}}

\newcommand\Mstarmin{M_\star^{\rm min}}
\newcommand\MDH{M_{\rm DH}}
\newcommand\MDHav{\left<M_{\rm DH}\right>}
\newcommand\MDHmin{M_{\rm DH}^{\rm min}}

\begin{document}

\title{LUMINOSITY DEPENDENT CLUSTERING OF \\ 
STAR-FORMING BzK GALAXIES AT REDSHIFT 2
\altaffilmark{1}
}

\author{
Masao Hayashi~\altaffilmark{2}, 
Kazuhiro Shimasaku~\altaffilmark{2},
Kentaro Motohara~\altaffilmark{3}, 
Makiko Yoshida~\altaffilmark{2},\\ 
Sadanori Okamura~\altaffilmark{2}        
and 
Nobunari Kashikawa~\altaffilmark{4}}

\email{hayashi@astron.s.u-tokyo.ac.jp}

\altaffiltext{1}{
Based on data collected at Subaru Telescope, which is operated by 
the National Astronomical Observatory of Japan.
Use of the UKIRT 3.8-m telescope for the observations is supported by NAOJ.}
\altaffiltext{2}{
Department of Astronomy, Graduate School of Science, 
University of Tokyo, Tokyo 113-0033, Japan}
\altaffiltext{3}{
Institute of astronomy, Graduate School of Science, University of
Tokyo, Mitaka, Tokyo 181-0015, Japan}
\altaffiltext{4}{
Optical and Infrared Astronomy Division, National Astronomical
Observatory, Mitaka, Tokyo 181-8588, Japan}

\begin{abstract}
We use the $BzK$ color selection proposed by \citet{dad04}
to obtain a sample of 1092 faint 
star-forming galaxies (hereafter sBzKs)
from 180 arcmin$^2$ in the Subaru Deep Field.
This sample represents star-forming galaxies at
$1.4 \lesssim z \lesssim 2.5$ down to $\KAB=23.2$, which
roughly corresponds to a stellar-mass limit of
$\simeq 1 \times 10^{10} M_\odot$.
We measure the angular correlation function (ACF) of these sBzKs
to be $w(\theta) = (0.58 \pm 0.13) \times \theta[\arcsec]^{-0.8}$
and translate the amplitude into the correlation length assuming
a reasonable redshift distribution.
The resulting value, $r_0 = 3.2^{+0.6}_{-0.7}$ $h^{-1}$ Mpc,
suggests that our sBzKs reside in haloes with a typical
mass of $2.8 \times 10^{11} M_\odot$.
Combining this halo mass estimate with those for brighter
samples of \citet{kon06}, we find that the mass of dark haloes
largely increases with $K$ brightness, a measure of the stellar mass;
the dark halo mass increases by as much as $10^{2-3}$
as $K$ brightness increases by only a factor of $\simeq 10$.
We also find that the halo occupation number, the number of galaxies
hosted in a dark halo, is higher for brighter sBzKs.
Comparison with other galaxy populations suggests
that faint sBzKs ($\KAB<23.2$) and Lyman Break Galaxies at $z \sim 2$ are
similar populations hosted by relatively low-mass haloes,
while bright sBzKs ($\KAB<21$) reside in haloes comparable to
or more massive than those of Distant Red Galaxies and Extremely Red 
Objects.
Using the extended Press-Schechter formalism,
we predict that present-day descendants of haloes hosting sBzKs
span a wide mass range depending on $K$ brightness,
from lower than that of the Milky Way up to those of richest clusters.
\end{abstract}

\keywords{cosmology: observations --- galaxies: evolution --- 
          galaxies: formation --- galaxies: high-redshift --- 
          galaxies: photometry}

\section{Introduction}

Galaxy mass is probably the most important parameter 
among those governing the evolution of galaxies.
Observations of the local universe show 
that more massive galaxies tend to have older stellar populations, 
lower star-formation activities, earlier morphological types, 
and higher metallicities \citep[e.g.,][]{gal05}.
In Cold Dark Matter (CDM) universes, dark haloes grow 
with time through merging of less massive haloes, 
implying that haloes with different masses have different 
evolutionary histories.
Observing galaxies over a wide mass range is thus 
crucial to place strong constraints on galaxy evolution.

It is extremely difficult to measure dark-halo masses 
of high-redshift galaxies. 
One can, however, {\it infer} them from galaxy clustering 
on the assumption that galaxies reside in dark haloes, 
since CDM models predict a monotonic correlation 
that more massive haloes are clustered more strongly 
\citep{mow02}.
Measuring galaxy clustering requires a large sample 
from a wide area.

Recent observations suggest that the era of $z \sim 2$ is 
important in galaxy evolution for various reasons; 
the cosmic star formation rate begins to drop at $z \sim 1$ -- 2 
from a flat plateau at higher redshifts \citep{dic03,fon03}; 
the morphological type mix of field galaxies changes remarkably 
at $z \sim 1$ -- 2 \citep{kaj01}; 
the number density of QSOs has a peak at $z \sim 2$ \citep{ric06}. 
Various color selection methods  
such as the Lyman-break technique \citep{guh90}  
have been proposed so far, in order to construct large samples 
of galaxies at high redshifts from imaging data.
However, color selections for galaxies at $z \sim 2$, 
the {\lq}redshift desert{\rq}, have not been 
as successful as those for other redshifts, 
mainly because of the lack of distinctive spectral features 
in the optical wavelength. 
 
Recently, a two-color selection with $B-z'$ and $z'-K$ has 
been proposed by \citet{dad04} to select  
galaxies at $1.4 \lesssim z \lesssim 2.5$ effectively 
with a low contamination.
Galaxies selected with this method, referred to as BzKs, 
make up a good approximation of a sample of all galaxies 
at $z \sim 2$ down to a limiting magnitude in the $K$ band,
roughly equal to a stellar mass limit.
Two types of BzKs, sBzKs and pBzKs, 
are defined according to the position 
in the $B-z'$ vs $z'-K$ space \citep{dad04}; 
{\lq}s{\rq} and {\lq}p{\rq} denote star-forming 
and passive, respectively.
It has been shown that most star-forming galaxies 
at $1.4 \lesssim z \lesssim 2.5$ meet the criteria for sBzKs 
irrespective of the amount of dust extinction \citep{dad04}.

Previous observations of BzKs have been limited 
primarily to bright objects with $K < 22$. 
Those studies have shown that 
bright sBzKs have star formation rates (SFRs) of 
$\sim 10^2 M_\odot$ yr$^{-1}$, 
stellar masses of $10^{11}M_\sun$, and 
dust extinctions of $E(B-V) \sim 0.4$ 
\citep[][hereafter K06]{dad04, dad05, kon06}.
Bright pBzKs are found to 
have stellar masses of $\gtrsim 10^{11}M_\odot$ 
with ages of $\sim1$ Gyr.
Clustering has also been examined recently 
using samples of bright BzKs from wide-field surveys 
(K06); 
both pBzKs and sBzKs are found to be clustered strongly. From 
these studies, bright BzKs are inferred to be 
progenitors of the present-day massive galaxies.
However, properties of faint, or low-mass, BzKs remain unrevealed
due to the difficulty in obtaining deep and wide-field NIR images.

\citet{ste04} have proposed another method 
to select galaxies at $z \sim 2$ \citep[see also][]{ade04}.
This method uses $U,G$ and $R$ photometry to select far-UV bright 
star-forming galaxies, referred to as BXs/BMs.
Clustering properties of BXs/BMs have been studied 
using large samples \citep{ade05}, 
since NIR imaging is not required.
Note, however, that this method will miss dusty galaxies and 
passive galaxies, both of which are faint in far-UV wavelengths.

We have been conducting a deep survey of BzKs 
in the Subaru Deep Field \citep{kas04,mai01}
to study properties of faint, or low-mass, BzKs.
Details of our survey and the photometric properties of 
BzKs detected from our survey, 
such as number densities, star-formation rates, 
and stellar masses, will be presented elsewhere 
(Motohara et al. in prep).

In this paper, we report results of clustering analysis 
of sBzKs detected in our initial data set over 180 arcmin$^2$.
Our study provides the first measurements of angular correlation 
of faint ($K<23.2$) sBzKs. 
Unfortunately, the number of pBzKs in the present sample 
is not large enough to obtain reliable measurements of their 
angular correlation.
Combining our result with those for bright ($K < 22$) 
sBzKs given in K06, 
we examine properties of dark haloes hosting sBzKs 
over a wide range of $K$ magnitude, or, equivalently, stellar mass.
Dark-halo masses are inferred from observed correlation lengths 
through an analytic model for the spatial clustering 
of dark haloes \citep[e.g.,][]{mow02}.
We also compare the dark-halo masses of sBzKs with 
those of other populations such as Lyman-break galaxies (LBGs).

The structure of this paper is as follows.
The optical and near-infrared data used to select BzKs 
are described in \S \ref{sec;data}.
In \S \ref{sec;bzk}, we construct a sample of faint 
sBzKs and check its consistency by comparing it 
with other samples.  
We measure the angular correlation function of sBzK in our sample 
in \S \ref{sec;analysis}. 
Then, for both our sample and K06's, 
the amplitude of the angular correlation function is 
transformed into the spatial correlation length.
Results and discussion are presented in \S \ref{sec;discussion}, 
and conclusions are given in \S \ref{sec;conclusion}.

Throughout this paper, magnitudes are in the AB system, 
and we adopt cosmological parameters of 
$\Omega_{m0}=0.3$, $\Omega_{\Lambda 0}=0.7$. and $\sigma_8 = 0.9$. 
We assume $h=0.7$ to compute the power spectrum of the cosmological 
density fluctuations and physical quantities except 
the correlation length, where $h$ is the Hubble constant 
in units of $100$ km s$^{-1}$ Mpc$^{-1}$. 
The correlation length is expressed in units of $h^{-1}$Mpc, 
since most of the previous studies adopted $h=1$.

\section{Data}
\label{sec;data}
\subsection{Optical Data}
The Subaru Deep Field (SDF) is a blank field centered on 
$(13^{\rm h} 24^{\rm m} 38\fs9, +27\arcdeg29\arcmin25\farcs9)$ (J2000).
The SDF has deep and wide-field Subaru/Suprime-Cam data of seven 
optical bandpasses, $B$, $V$, $R$, $i'$, $z'$, NB816, and NB921, 
obtained for the Subaru Deep Field Project 
\citep{kas04}.
In this paper, we combine the $B$ and $z'$ data 
with the $K$ data described below, to select BzK galaxies.
The exposure time and the $3\sigma$ limiting magnitude 
on a $2\arcsec$ aperture are 595 min and 28.45 mag for $B$ and 
504 min and 26.62 mag for $z'$, respectively.
Both images have been convolved to a seeing size of $1\farcs14$ (FWHM), 
the value for the final $J$ and $K$ images (see below), 
and have an identical sky coverage of 
$29\farcm7 \times 36\farcm7 = 1090$ arcmin$^2$ 
with a pixel scale of $0\farcs202$ pixel$^{-1}$.
\begin{figure}[h]
 \begin{center}
  \plotone{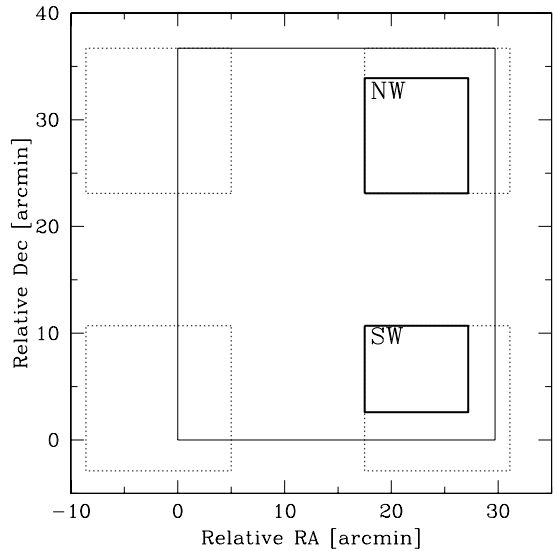}
  \caption{The sky area of the SDF observed with 
           Suprime-Cam and WFCAM.
           The thin solid lines outline the $29\farcm7 \times 36\farcm7$ 
           area imaged with Suprime-Cam ($B,z'$).
           The four dotted rectangles correspond to the four 
           fields with WFCAM $J$ and $K$ data. 
           we used in this study only the two subregions enclosed with 
           thick solid lines in the northwest and southwest fields.
           }
  \label{SDF}
 \end{center}
\end{figure}

\subsection{Near-Infrared Data}
We observed the SDF in the $J$ and $K$ bands 
with the wide-field camera (WFCAM) on the United Kingdom Infra-Red 
Telescope (UKIRT) on 2005 April 14 -- 15. 
WFCAM is composed of four 2048$\times$2048-pixel 
detectors with a large spacing of 12.3 arcmin. 
Each detector covers $\simeq 13\arcmin\times13\arcmin$ 
of sky with a pixel scale of $0\farcs4$. 
We made only a single pointing for each passband 
to go significantly deeper than the previous wide-field $K$ data 
used to study BzK galaxies.

Figure \ref{SDF} shows the sky area of the SDF 
imaged with Suprime-Cam and WFCAM.
The Suprime-Cam $29\farcm7 \times 36\farcm7$ field 
is outlined by the thin solid line.
The four dotted rectangles correspond 
to the regions imaged with the four detectors of WFCAM.
The large spacing between the detectors prevented us from 
covering the Suprime-Cam field effectively. 
The total area overlapping with the Suprime-Cam field 
amounts, however, to 410 arcmin$^2$, large enough 
to measure the angular clustering of faint BzK galaxies.

The exposure time was 150 min for $J$ and 294 min for $K$. 
The variations in sensitivity and PSF 
among the four detectors were found to be small.
The seeing size in the final images varied from $1\farcs04$ 
to $1\farcs14$ in $K$.

Data reduction was made in the standard manner for NIR imaging data, 
but special care was taken to exclude spurious objects 
due to crosstalk of bright objects.
All the images, except the one with the worst seeing size, 
were smoothed with a Gaussian kernel 
so that their PSF sizes be $1\farcs14$ FWHM.
The 2MASS catalog \citep{skr06} was used to conduct astrometry 
and to derive the magnitude zero point.

We measure the $5\sigma$ limiting magnitude on a $2\arcsec$-diameter 
aperture for each detector. 
The shallowest (brightest) limiting magnitude 
among the four detectors is 23.3 mag in $J$ and 23.5 mag in $K$.
The variation in limiting magnitude among the four detectors 
is less than 0.2 mag for both bandpasses.
Details of our observation and data reduction will be 
described elsewhere (Motohara et al. in prep).

\subsection{$K$-Selected Catalog}
Object detection and photometry are made using 
SExtractor 2.1.3 \citep{ber96}. 
The $K$-band is chosen as the detection band.
For each object detected in the $K$ image, 
$2\arcsec$-diameter magnitude is measured for $B$ and $z'$ 
using double-image mode of SExtractor to derive colors.
If an object is undetected (fainter than the $2\sigma$ magnitude) 
in $B$ or $z'$, its magnitude is replaced 
with the $2\sigma$ magnitude.
We limit our catalog to $K=23.5$, 
the shallowest $5\sigma$ limiting magnitude 
among the four detectors. 
We have in total 8884 objects in the overlapping fields 
(see Figure \ref{SDF}).
The catalog thus constructed is then corrected for 
Galactic extinction using the dust map of \citet{sch98}: 
$A(B) = 0.07$ mag, $A(z') = 0.02$ mag, and $A(K) = 0.01$ mag. 

\subsection{Detection Uniformity}
Clustering analysis requires high uniformity in object detection 
over the image for all passbands, 
since a variation in detection completeness may 
produce spurious clustering signals.
While the WFCAM $J$ and $K$ images are very uniform, 
sky noise at the four edges of the Suprime-Cam images is 
systematically large and non-uniform
because of dithered pointing with large angles.

We examine detection uniformity for each of the four overlapping 
fields by dividing the field into a few dozen of sub-areas 
and estimating the sky noise in each sub-area for 
$B$, $z'$, and $K$.
On the basis of these noise maps, we find that the two subregions 
surrounded by thick solid lines in Figure \ref{SDF}
have a good detection uniformity in all three passbands.
These subregions are well within the good $S/N$ area 
defined in \citet{kas04}.
In what follows we call these two subregions 
the NW and SW subregions.
The area of NW and SW subregions are 100 arcmin$^2$ and
80 arcmin$^2$, respectively.

In this paper, we examine angular clustering 
only in the NW and SW subregions.
Among the 8884 objects found in the four overlapping fields, 
3961 are included in these two subregions.
The two eastern fields have also sub-areas with 
a good detection uniformity, 
but they are all too small for clustering analysis.

\section{BzK Galaxies in the Subaru Deep Field}
\label{sec;bzk}

\subsection{Selection of BzK Galaxies}
\label{subsec;selection_bzk}
In the $B-z'$ vs $z'-K$ plane, sBzKs are defined to be objects 
with $(z'-K)-(B-z')>-0.2$, while the criteria for pBzKs 
are defined as $(z'-K)-(B-z')<-0.2 \ \cap \ z-K>2.5$ 
\citep{dad04}.
Stars and foreground galaxies are separated well 
from BzK galaxies in this plane \citep{dad04}.
We also regard objects with $K<18 \ \cap$ CLASS\_STAR $> 0.9$ as stars, 
where CLASS\_STAR is the stellarity index parameter from SExtractor. 
Although the validity of the $BzK$ selection for objects 
fainter than $K \simeq 22$ has not been examined well, 
we assume that the selection and the classification 
into two types are applicable to our objects as well.

In the selection of BzKs, special care is required for 
objects undetected in $B$ or $z'$, 
i.e., $B>\Btwosigma$ or $z'>\ztwosigma$, 
where $\Btwosigma$ and $\ztwosigma$ are 
the $2\sigma$ limiting magnitudes.
We treat these objects as follows:

\begin{itemize}
\item Objects undetected in $B$ but detected in $z'$. \\
      If their $\Btwosigma - z'$ and $z'-K$ colors 
      meet the pBzK criteria, they are classified as pBzKs, 
      since their true $B-z'$ colors, which are 
      redder than $\Btwosigma - z'$, also meet $(z'-K)-(B-z')<-0.2$.
      On the other hand, classification is not applied 
      to those whose $\Btwosigma - z'$ and $z'-K$ colors 
      satisfy the sBzK criterion, since their true 
      $B-z'$ colors may be too red to be classified as sBzKs.
\item Objects detected in $B$ but undetected in $z'$. \\
      If their $B - \ztwosigma$ and $\ztwosigma - K$ 
      colors meet the sBzK criterion, they are classified as sBzKs, 
      since they do not go out of the sBzK region even when 
      true $z'$ magnitudes, which are fainter 
      than $\ztwosigma$, are used instead.
      On the other hand, classification is not applied 
      to those with $B - \ztwosigma$ and $\ztwosigma - K$ 
      satisfying the pBzK criteria, since their true 
      $B-z'$ may be too blue to be classified as pBzKs.
\item Objects undetected in both $B$ and $z'$. \\
      These objects are not classified, because their 
      $B-z'$ colors can take any values.
\end{itemize}

\begin{figure}[h]
 \begin{center}
  \plotone{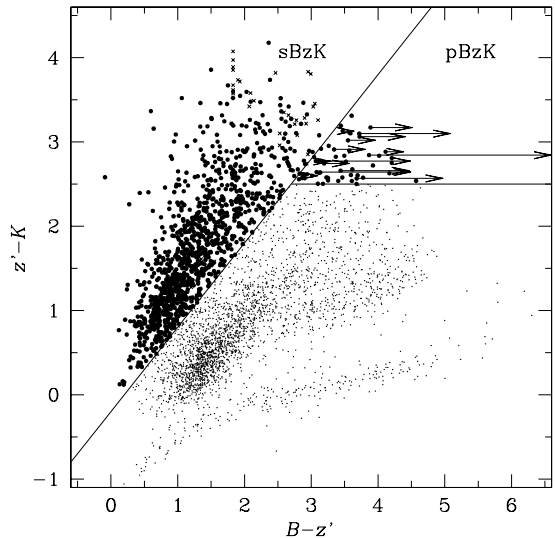}
  \caption{$B-z'$ vs $z'-K$ distribution of all objects 
           in the NW and SW subregions.
           For objects undetected in $B$ or $z'$, 
           $\Btwosigma$ and $\ztwosigma$
           have been used, respectively, in the plot.
           Filled circles represent BzKs; 
           those with a horizontal arrow are undetected in $B$. 
           Crosses represent unclassified objects.
           }
  \label{BzK}
 \end{center}
\end{figure}

The numbers of sBzKs and pBzKs selected 
are 1092 and 56, respectively (Table \ref{N}).
Figure \ref{BzK} shows the $B-z'$ vs $z'-K$ distribution 
of all objects in the NW and SW subregions.
\begin{table}
 \caption{Numbers of sBzKs, pBzKs and unclassified objects}
\begin{center}
\begin{tabular}{ccccc}
\hline\hline
\multicolumn{2}{c}{Detection$^*$}& sBzK& pBzK& unclassified\\
$B$ & $z'$ & & & \\ 
\hline
$\circ$ & $\circ$ & 1092 & 40 & 0 \\
$\times$ & $\circ$ & 0 & 16 & 20 \\
$\circ$ & $\times$ & 0 & 0 & 0 \\
$\times$ & $\times$ & 0 & 0 & 9 \\
\hline
\multicolumn{2}{c}{total} & 1092 & 56 & 29 \\
\hline\hline
\label{N}
\end{tabular}
\end{center}
\vspace{-0.7cm}
* $\circ$ and $\times$ represent detection and non-detection,
respectively.
\end{table}

Our $B$, $z'$, and $K$ bandpasses are not exactly the same as 
those originally used to define the BzK selection criteria 
in \citet{dad04}.
We use \citet{gun83}'s stellar spectrophotometric atlas 
to find that the $B-z'$ colors of stars defined in our system 
can be bluer than those in \citet{dad04} up to 0.2 mag, 
depending on the spectral type, 
while there is little difference in $z'-K$.
We also make a similar evaluation using the spectral templates 
of local galaxies (E to Im) given in \citet{col80} 
redshifted to $z \sim 2$.
It is found that the $B-z'$ colors of these galaxies 
in our system are at most 0.1 mag bluer than those in \citet{dad04} 
while our $z'-K$ colors are at most 0.1 mag redder.

These results imply that applying the original sBzK boundary, 
$(z'-K) - (B-z') > -0.2$, to our data will select additional 
objects near the boundary which, if measured in the system of
\citet{dad04}, may not be selected.
Our sBzK sample will be reduced by $\simeq 15$ \% 
when the selection boundary is tightened by $0.2$ mag 
(i.e., $(z'-K) - (B-z') > 0.0$).
This reduction should be an upper limit to the effect of the bandpass 
differences, since the net offset in the $B-z'$ vs $z'-K$ 
plane of sBzKs due to the bandpass differences 
will be on average more modest than $0.2$ mag.
Because this reduced sample gives a very similar ACF, 
we have decided not to adjust the selection boundary in our study. 

We find that the $2\arcsec$-aperture $K$ magnitudes 
of our BzKs are on average 0.3 mag fainter 
than the total magnitudes (we adopt MAG\_AUTO magnitudes 
for total magnitudes).
For this reason, we regard that our BzK sample is 
flux-limited to a total magnitude of $K=23.2 (=23.5-0.3)$.

\subsection{Number Counts}
\begin{figure}[h]
 \begin{center}
  \plotone{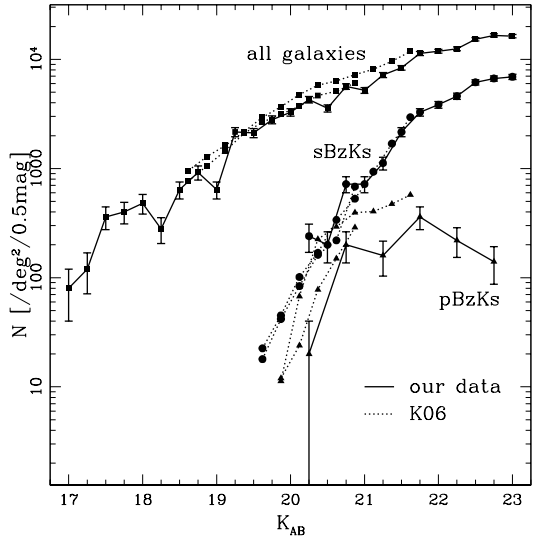}
  \caption{Number counts of sBzKs, pBzKs, and all galaxies 
           as a function of total magnitude (MAG\_AUTO).
           The circles, triangles, and squares represent 
           sBzKs, pBzKs, and all galaxies, respectively.
           Our data are connected by the solid line, while 
           K06's are connected by the dotted line. 
           The K06 data are plotted separately for 
           the two survey fields.
           }
  \label{numbercount}
 \end{center}
\end{figure}

Figure \ref{numbercount} compares the number counts of sBzKs, 
pBzKs, and all galaxies from our data with those given by K06.
Since K06's data are from two wide fields, 
EIS Deep3A field and Daddi field, we plot the counts 
from these two fields separately.
It is found that our number counts of sBzKs and all galaxies 
are consistent with those of K06 
while pBzK are lower by a factor of 2 - 3.
This disagreement in the number counts of pBzK may be partly due to cosmic 
variance. 
Indeed, even in the two fields surveyed by K06, both of which are 
several times larger than our total area, the counts are 
different significantly.
A small part of the disagreement may also result 
from our special treatment in the $BzK$ selection 
for objects undetected in $B$ or $z'$ 
described in Section \ref{subsec;selection_bzk}, 
which has not been considered in K06. 

\section{Clustering Analysis}
\label{sec;analysis}

\subsection{Angular Correlation}
We derive the angular correlation function (ACF) 
separately for the sBzKs and pBzKs.
We do not, however, show the results of the pBzKs, 
since reliable measurements are not obtained   
due to their small sample size (56 objects).

\begin{figure}[th]
  \begin{center}
  \plotone{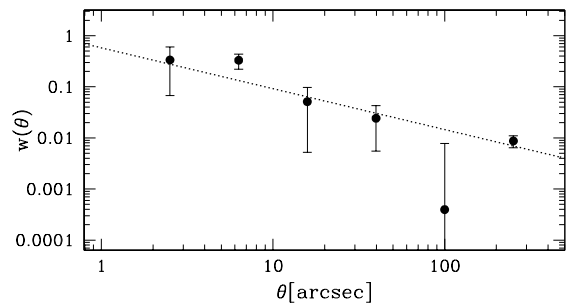}
  \caption{The angular correlation function of sBzKs.
  The error bars correspond to the 1$\sigma$ errors 
  obtained from bootstrap resampling.
  The dotted line is the best-fit power law of $A \theta^{-0.8}$.
  }
  \label{ACF}
  \end{center}
\end{figure}

We first measure the ACF of sBzKs for the NW and SW subregions 
independently with angular bins of $\log \theta[\arcsec] = 0.4$, 
using the estimator given in \citet{lan93}. 
Random data with 50000 points are used.
Errors in the measurements are estimated by 
the bootstrap resampling method \citep{lin86, mo92}.
Then the measurements for each subregion are corrected 
for the integral constraint, on the assumption that the true ACF 
obeys a power law of $ w(\theta) \propto \theta^{-0.8}$.
Finally, the measurements from the two subregions are averaged 
to obtain the ACF of sBzKs in the SDF 
(Figure \ref{ACF}).

The amplitude of the ACF, $A$, is obtained from a fit of 
$w(\theta) = A\theta^{-0.8}$ to the data, 
and is given in Table \ref{count_etc}.
Our sBzKs have $A = 0.58\pm0.13$ [arcsec$^{0.8}$], 
where the errors correspond to the range 
in which the increase in $\chi^2$ from the best-fit value 
is less than unity.
Our ACF measurement for sBzKs could suffer from cosmic variance, 
since each of the two sub-fields is only about 10 Mpc a side 
at $z=2$. 
A survey of a larger area is clearly required to obtain results 
robust against possible variance. 
However, we infer that the amplitude of the cosmic variance 
in our sample will not be intolerably large, 
since the ACFs from the two sub-fields agree within the statistical
errors, and since the number counts of our sBzKs are consistent 
with those of K06.  

\subsection{Bright Samples by K06}
\begin{figure}[h]
  \begin{center}
  \plotone{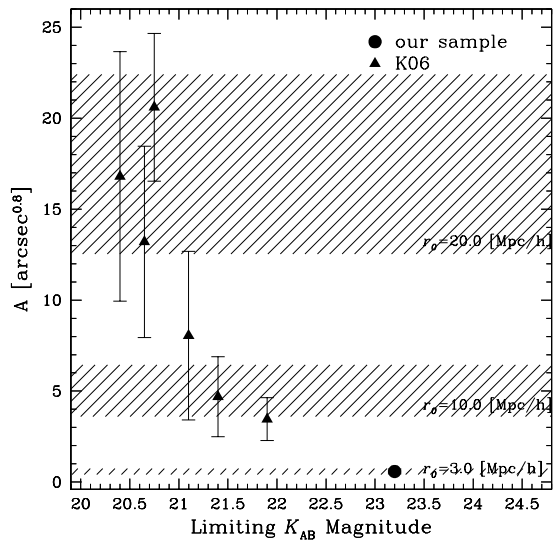} 
  \caption{Amplitude of the ACF, $A$, as a function of 
   limiting $K$ total magnitude.    
   The filled circle shows the measurement for our sample, 
   while the filled triangles correspond to the six samples 
   with different limiting magnitudes over $K=20.4$ and $21.9$ 
   given in K06.
   The amplitudes for $r_0 = 20, 10$ and $3 h^{-1}$Mpc 
   calculated using the adopted redshift distribution for sBzKs 
   (Gaussian with $z_c = 1.9$ and $\sigma_z = 0.35$) are 
   shown by the shaded regions, whose width
   corresponds to the adopted uncertainty 
   in $\sigma_z$ of $\pm 0.1$.
   See subsection \ref{subsec;nz} for details. 
   }
  \label{Aw_K}
  \end{center}
\end{figure}

Figure \ref{Aw_K} shows the observed ACF amplitude against limiting
$K$ magnitude. 
The filled circle represents the amplitude of our sBzKs, 
while the six filled triangles correspond to 
the amplitudes for the six samples of K06 
with different limiting magnitudes over $K=20.4$ and $21.9$. 
The ACF amplitude of sBzKs is found to 
decrease rapidly with decreasing $K$ brightness.
The brightest samples of K06 have 
$A \simeq 15$ -- 20 arcsec$^{0.8}$, which is about 30 times 
higher than ours.

In the next subsection, the spatial correlation lengths, $r_0$, 
are calculated from the ACF amplitudes 
to discuss the clustering of sBzKs in real space.
For the reader's reference, 
the three shaded regions in Figure \ref{Aw_K} indicate, 
respectively, the expected ACF amplitude for 
$r_0$ = 3, 10 and 20 [$h^{-1}{\rm Mpc}$] calculated 
using the redshift distribution for sBzKs adopted 
in the next subsection 
(Gaussian with $z_c = 1.9$ and $\sigma_z = 0.35$); 
for each $r_0$, the spread in amplitude
corresponds to the adopted uncertainty in the $\sigma_z$ of $\pm 0.1$.

\subsection{Spatial Correlation Length of sBzKs}
The correlation length, $r_0$, for sBzKs 
can be calculated from the amplitude of the ACF, $A$, 
using Limber's equation \citep{lim53,pee80} 
if their redshift distribution is known.

\subsubsection{Redshift Distribution of sBzKs} 
\label{subsec;nz}
The redshift distribution, $N(z)$, of sBzKs has not been 
established even for bright objects with $K<22$  
owing to the lack of a large, flux-limited spectroscopic sample.
In this study, we assume that the $N(z)$ of sBzKs 
is not dependent on $K$ magnitude, 
and adopt a Gaussian distribution as a simple approximation. 
We determine the central redshift, $z_c$, and the standard 
deviation, $\sigma_z$, of the Gaussian as follows.

First, we use a preliminary result of the $N(z)$ 
of sBzKs (81 objects, most of which are $K=21-22$) obtained from 
a $K$-selected spectroscopic survey in the EIS Deep3A field 
with VLT/VIMOS down to $K = 22.0$ 
(E. Daddi, private communication) 
to determine $z_c$ and infer $\sigma_z$.
This is probably the most reliable measurement of the $N(z)$ 
of sBzKs at present because it is based on 
the largest spectroscopic sample.
We fit a Gaussian to this $N(z)$ after removing 
a strong spike at $z \sim 1.5$ (which could be due to large-scale 
structure) and limiting the redshift range to $1.0<z<3.0$, 
and obtain $z_c$ = 1.9 and $\sigma_z$ = 0.4. From this result, 
we adopt $z_c$ = 1.9 for our Gaussian distribution.
Note that the Limber transformation is insensitive to the change 
in $z_c$; for example, changing $z_c$ by $\pm 0.2$ around 1.9 
changes $r_0$ only by 2.5 \%. 

\begin{figure}[ht]
  \begin{center}
  \plotone{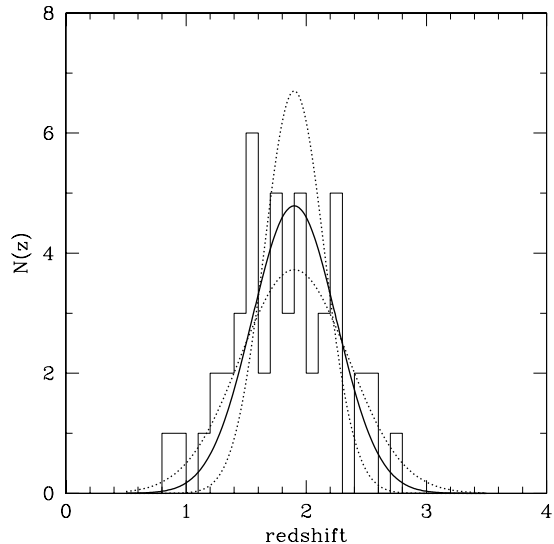}
  \caption{
   The redshift distribution of sBzKs. 
   The histogram represents the distribution of 28 sBzKs 
   with a spectroscopic redshift given in \citet{dad04}. 
   The solid curve indicates a Gaussian distribution function 
   with $z_c = 1.9$ and $\sigma_z = 0.35$ 
   used to calculate the correlation length.
   The two Gaussians with $\sigma_z = 0.25$ and $0.45$ 
   are plotted by the dotted line to show the adopted uncertainty 
   in the redshift distribution.
   The amplitude of the Gaussians has been normalized 
   so that their areas be equal to the area of 
   the histogram.
   }
  \label{selection_func}
  \end{center}
\end{figure}

Next, we determine $\sigma_z$ and express the uncertainty 
in our Gaussian distribution by the range of $\sigma_z$. 
We adopt $\sigma_z = 0.35 \pm 0.1$ for the following reasons.
We find that, for a fixed $A$ value, 
the range of $r_0$ corresponding to 
the range $0.25 < \sigma_z < 0.45$ covers not only the $r_0$ 
value calculated using the raw $N(z)$ from the EIS Deep3A field, 
but also from those given in \citet{dad04} and \citet{red05}. 
Here, the data of \citet{red05} are another measurement of the redshift 
distribution based on a large sample, although the sample is 
not $K$-limited but optically selected and thus could be 
biased toward UV bright objects.
We also find that the best-fit Gaussian parameters 
for the redshift distribution data of 
\citet{dad04} and \citet{red05} are similar to 
the values we adopt above. 
Figure \ref{selection_func} compares the adopted Gaussian 
function with the $N(z)$ of \citet{dad04}.

It is true that a Gaussian distribution is not an excellent 
approximation to the existing $N(z)$ data, 
especially for the preliminary result of Daddi et al. 
(private communication) with a large spike.
However, it will be reasonable to expect that the true $N(z)$ 
derived from a sample large enough to smooth out statistical 
noise and cosmic variance is not so far from a Gaussian centered 
at $z \sim 2$, 
since the isolation of $1.4<z<2.5$ galaxies from others 
using the BzK selection will not be perfect. 
It may be worth noting that Lyman-break galaxies, which are also 
selected in a two-color plane, are known to have 
a Gaussian-like distribution \citep{yos06}.
Our purpose is not to infer the true distribution 
but to find a reasonable expression of it 
which can be used in the Limber transformation. 
In this sense, 
the adopted Gaussian and uncertainty, $z_c = 1.9$ and 
$\sigma = 0.35 \pm 0.1$, are a reasonable parameterization, since 
they include all the variations in the existing redshift distribution
measurements in terms of resultant $r_0$ values.
It is also found that the $r_0$ value calculated using 
a top-hat redshift distribution of $1.4 < z < 2.5$ (ideal 
selection function for BzKs) falls within the $r_0$ range 
defined by the adopted uncertainty in the Gaussian.

Finally, we should note that our assumption that the $N(z)$ 
does not depend on $K$ magnitude must be tested 
using a future spectroscopic follow-up, 
although the observation of \citet{red05} may support this assumption. 
They derived $N(z)$ for $K < 21.8$ and $K > 21.8$ 
(with a limiting magnitude of $K = 24.3$) separately, 
which are found not to be largely different.

\subsubsection{Spatial Correlation Length}
The correlation length of our sBzKs derived in this manner 
is $r_0 = 3.2^{+0.6}_{-0.7}$ [$h^{-1}$Mpc] in comoving 
units (see also Table \ref{count_etc}).
The errors in $r_0$ are calculated by summing up in quadrature
the errors from the measurement of $A$ and the errors due to 
the adopted uncertainty in $\sigma_z$ ($\pm 0.1$).

\begin{figure}[h]
  \begin{center}
  \plotone{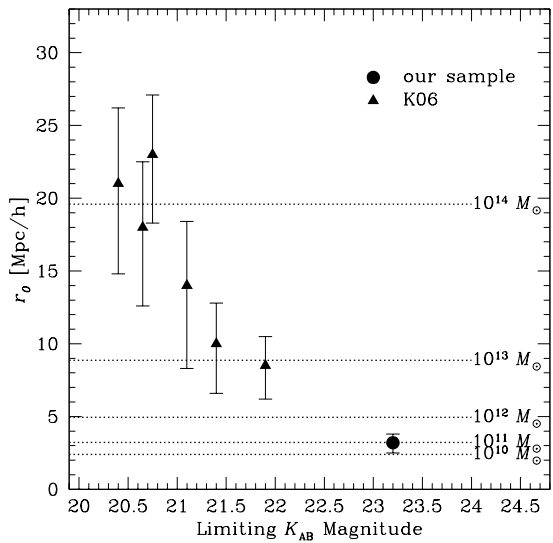} 
  \caption{Correlation length, $r_0$, as a function of 
   limiting $K$ total magnitude. 
   The filled circle shows the measurement for our sample, 
   while the filled triangles correspond to the six samples 
   with different limiting magnitudes over $K=20.4$ and $21.9$ 
   given in K06. 
   The dotted horizontal lines indicate
   the minimum mass of dark haloes inferred from $r_0$.
   }
  \label{r0_K}
  \end{center}
\end{figure}

Figure \ref{r0_K} shows the correlation lengths for 
our sBzK sample and the six samples of K06.\footnotemark[1]
\footnotetext[1]{
The errors in $r_0$ of the samples of K06 are not 
very large in spite of the small sample sizes, 
because brighter sBzKs are clustered more strongly 
and thus have stronger signals.
This in turn means that larger samples are required 
to measure the angular correlation function 
for fainter objects with the same uncertainty.
}
It is found from Figure \ref{r0_K} that $r_0$ increases 
rapidly with increasing $K$ brightness.
The brightest sample of K06 has 
$r_0 \simeq 20$ $h^{-1}$Mpc, which is six times larger than 
that of ours.

\begin{table*}
 \caption{Numbers, ACF amplitudes, correlation lengths, 
 and dark-halo masses of sBzKs in our sample and K06's}
\begin{center}
 \tabcolsep 3pt
 \begin{tabular}{cccccccc}

     \tableline
     \tableline
                  & Field &  $K_{\rm lim}$  & N & A[arcsec$^{0.8}$]&
  $r_0$[$h^{-1}$Mpc] & $\MDHav$[$M_\odot$] & 
                       $\MDHmin$[$M_\odot$] \\  
     \tableline
     this work     & SDF      &  23.2 & 1092 & 0.58$\pm$0.13 &
  $3.2^{+0.6}_{-0.7}$ & $2.8^{+4.2}_{-2.3} \times 10^{11}$ &
    $1.0^{+1.8}_{-0.9} \times 10^{11}$  \\
     K06 & Daddi-F  &  20.4 & 21   & 16.8$\pm$6.86 &
  $21^{+5.2}_{-6.2}$  & $1.6^{+0.9}_{-0.9} \times 10^{14}$ &
    $1.1^{+0.9}_{-0.6} \times 10^{14}$  \\
                   &          &  20.7 & 43   & 13.2$\pm$5.26 &
  $18^{+4.5}_{-5.4}$  & $1.1^{+0.8}_{-0.6} \times 10^{14}$ &
    $8.0^{+6.0}_{-5.0} \times 10^{13}$  \\
                   &          &  21.1 & 92   & 8.05$\pm$4.63 &
  $14^{+4.4}_{-5.7}$  & $6.4^{+4.6}_{-4.8} \times 10^{13}$ &
    $4.2^{+4.8}_{-3.4} \times 10^{13}$  \\
                   & Deep3A-F &  20.7 & 27   & 20.6$\pm$4.06 &
  $23^{+4.1}_{-4.7}$  & $1.9^{+0.8}_{-0.8} \times 10^{14}$ &
    $1.5^{+0.6}_{-0.6} \times 10^{14}$  \\
                   &          &  21.4 & 129   &  4.69$\pm$2.20 &
  $10^{+2.8}_{-3.4}$  & $2.8^{+2.4}_{-2.0} \times 10^{13}$ &
    $1.6^{+1.7}_{-1.3} \times 10^{13}$  \\
                   &          &  21.9 & 387   &  3.46$\pm$1.18 &
  $8.5^{+2.0}_{-2.3}$  & $1.8^{+1.4}_{-1.2} \times 10^{13}$ &
    $9.0^{+9.0}_{-6.3} \times 10^{12}$  \\
     \tableline
     \tableline
 \end{tabular}
 \label{count_etc}
\end{center}
\end{table*}

\section{Discussion}
\label{sec;discussion}

\subsection{Masses of Dark Haloes Hosting sBzKs}

The standard CDM model predicts that at any redshifts 
more massive dark haloes are on average 
more clustered \citep[e.g.,][]{mow02}. 
We use the clustering strengths of our sBzK sample 
and those of K06 to infer the masses of their hosting haloes 
in the following manner.
First, for an sBzK sample with a given limiting magnitude, 
we derive the amplitude of the spatial correlation 
function at 8 $h^{-1}$ Mpc, $\xisBzK(8 h^{-1} {\rm Mpc})$, 
from the observed $r_0$ assuming $\xisBzK \propto r^{-1.8}$. 
We then calculate the effective bias parameter, $\beff$, 
from 
\begin{equation}
\beff = \sqrt{ \xisBzK(8 h^{-1} {\rm Mpc})
                 \over{\xiDM(8 h^{-1} {\rm Mpc})} }, 
\end{equation}
where $\xiDM$ is the predicted correlation function of 
dark matter at $z=2$.
Finally, we use the analytic formula of dark-halo biasing 
given in \citet{she01}
to obtain the mass of dark haloes hosting the sBzKs.

We consider two definitions of halo mass, 
(i) average mass and (ii) minimum mass.
Average mass, $\MDHav$, is defined by: 
\begin{equation}
\beff = b(\MDHav), 
\end{equation}
where $b(\MDH)$ is the bias parameter of haloes with mass $\MDH$.
We regard $\MDHav$ as the typical halo mass of the sample.
Minimum mass, $\MDHmin$, is defined through: 
\begin{equation}
\beff = { \int_{\MDHmin}^{\infty} b(\MDH) n(\MDH) d\MDH
        \over{ \int_{\MDHmin}^{\infty} n(\MDH) d\MDH } }, 
\end{equation}
where $n(\MDH)d\MDH$ is the mass function of dark haloes.
In this definition, $\MDHmin$ corresponds to the mass of haloes 
hosting the faintest galaxies in the sample 
(In this paper we assume that bias increases monotonically 
with galaxy brightness).

Our definition of $\MDHmin$ can be regarded as a simplification 
of the halo model approach which uses three parameters 
including $\MDHmin$ to model galaxy clustering
\citep{baw02,bul02,mas02,ham04}; 
the other two parameters are used to model 
the mass dependence of the number of galaxies per halo.
In this study, we do not adopt this approach, 
but place a constraint only on $\MDHmin$ for simplicity, 
since this approach requires an accurately measured ACF shape 
with small angular bin sizes.

The average dark-halo mass of our sBzK sample is estimated 
to be $\MDHav \simeq 3 \times 10^{11} M_\odot$. 
In contrast, the sBzKs of K06 samples are 
found to be hosted by massive haloes with 
$\MDHav \simeq 2 \times 10^{13}$ -- $2 \times 10^{14} M_\odot$.

The difference between $\MDHav$ and $\MDHmin$ is less than 
factor 3.
This is because fainter objects are more numerous, 
thus contributing more to the overall clustering of the sample.
For the readers' reference, 
The dotted horizontal lines of Figure \ref{r0_K} 
indicate $\MDHmin$ as a function of $r_0$.

These results show that sBzKs reside in dark haloes 
with a mass range of as wide as three orders of magnitude, 
$10^{11-14} M_\odot$, with fainter objects being found in 
less massive haloes.
Typical haloes hosting faint sBzKs ($K<23.2$) are 
several times less massive than that of the Milky Way 
($\MDH \simeq (1-2) \times 10^{12} M_\odot$: 
e.g., \citet{sak03}) 
while those hosting bright sBzKs 
with limiting magnitudes of $K=20.4$ to $21.9$ 
have masses comparable to those of present-day galaxy groups and clusters.

\subsection{Ratio of Stellar Mass to Dark Halo Mass}
We estimate the minimum stellar mass, 
which corresponds to that of the faintest galaxies in the sample,
for our sample and the six samples of K06 
from the $K$ magnitude and $z'-K$ color 
of the faintest object in each sample, 
using equations (6) and (7) given in \citet{dad04}.
These equations were derived using bright sBzKs which have accurate 
stellar masses based on a fit of model spectra to multicolor data.
\citet{dad04} found the typical error in mass estimates 
from these equations to be $\Delta \log M_\star=0.2$.

\begin{figure}[h]
  \begin{center}
  \plotone{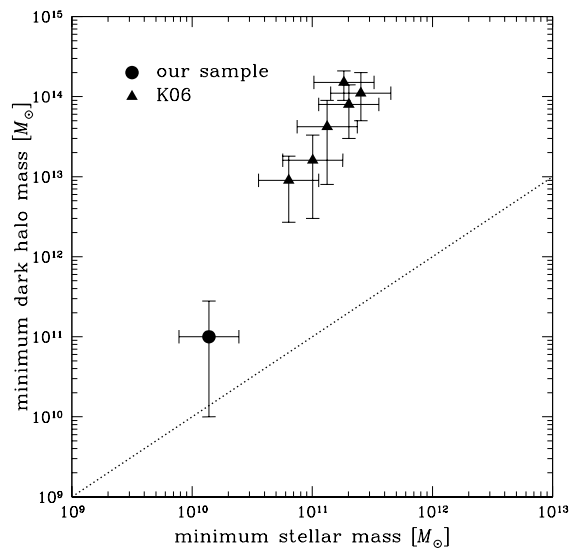}
  \caption{Minimum dark-halo mass plotted against 
   minimum stellar mass for sBzKs.
   The filled circle indicates our sample, 
   while the filled triangles correspond to the six samples 
   with different limiting magnitudes given in K06.  
   Minimum stellar masses correspond 
   to the limiting magnitudes of the samples.}
  \label{stellar_mass}
  \end{center}
\end{figure}

Figure \ref{stellar_mass} plots the minimum stellar mass of 
sBzKs against the minimum mass of their hosting haloes.
The minimum stellar mass of our sample is 
$\Mstarmin \simeq 1 \times 10^{10} M_\odot$, 
while those of K06 are three to ten times higher.
On the other hand, the minimum dark-halo mass of our sample 
is $\MDHmin \simeq 1 \times 10^{11} M_\odot$, while those of 
K06 are two to three orders of magnitude higher.
It is thus concluded that dark-halo mass increases more rapidly 
than the stellar mass of individual sBzKs residing in the halo.

We note that the stellar-mass estimation for our sample 
is less reliable than that for K06.
Equations (6) and (7) in \citet{dad04} 
have not been tested for faint sBzKs 
like those in our sample, which may have systematically 
different stellar populations and dust extinctions.
The observed trend of increasing $\MDHmin/\Mstarmin$ with 
$\Mstarmin$ will disappear if the $\Mstarmin$ of our sample 
is overestimated by more than factor ten, 
although such a large overestimation seems to be unlikely.

\subsection{Halo Occupation Number}
\label{sec;occupation}
The halo occupation number of a given class of galaxies 
is defined as the average number of this class of galaxies 
hosted in a single dark halo.
We estimate the halo occupation number of 
sBzKs for two magnitude limits, $K<23.2$ and $K<21.9$, 
using our sample and the faintest sample of K06, 
and then examine whether or not occupation number depends 
on luminosity.

The halo occupation number for sBzKs with $K=23.2$ 
is estimated by dividing the number density of 
our sBzKs by that of dark haloes hosting them, 
i.e., haloes more massive than $\MDHmin$.
We have 1092 objects in a survey volume of 
$7.1 \times 10^5 \rm{Mpc^3}$.
This leads to a number density of $2.4 \times 10^{-3}$ Mpc$^{-3}$
after correction for detection completeness of a modest amount.
The number density of dark haloes is calculated to be 
$1.7 \times 10^{-2} \rm{Mpc^{-3}}$ 
from the analytic formula of the halo mass function 
given by \citet{she01}.
Thus, the halo occupation number of sBzKs with $K<23.2$ 
is estimated to be $\simeq 0.1$.

Similarly, we use the faintest sample of K06 with 
$K<21.9$ to estimate the halo occupation number to be $\simeq 4$.\footnotemark[2]
\footnotetext[2]{
We do not use the brighter five samples of K06, 
because sBzKs in them are hosted by more massive 
haloes, whose number densities, and thus occupation numbers, 
have increasingly large errors owing to a steep slope 
of the mass function.}
Here, the number of sBzKs in this sample is 387, 
the survey volume is $3.2 \times 10^{-4} \rm{Mpc^{-3}}$, 
and the number density of the hosting haloes 
is $7.5 \times 10^{-5} \rm{Mpc^{-3}}$.

These two estimates show that the occupation number of sBzKs 
increases strongly with increasing halo mass.
Haloes with $\MDH \gtrsim 10^{13} M_\odot$ have several sBzKs 
on average, while only one tenth (on average) of haloes with 
$\MDH \gtrsim 10^{11} M_\odot$ host an sBzK.
It is reasonable that haloes with $\gtrsim 10^{13} M_\odot$ 
host multiple sBzKs, since they are as massive as present-day 
groups and clusters of galaxies.
On the other hand, the very small occupation number 
for haloes with $\gtrsim 10^{11} M_\odot$ may 
imply that they are not massive enough 
to always feed a galaxy more luminous than $K=23.2$.

\subsection{Comparison of $r_0$ between sBzKs and Other Objects}

\begin{figure}[!h]
  \begin{center}
  \plotone{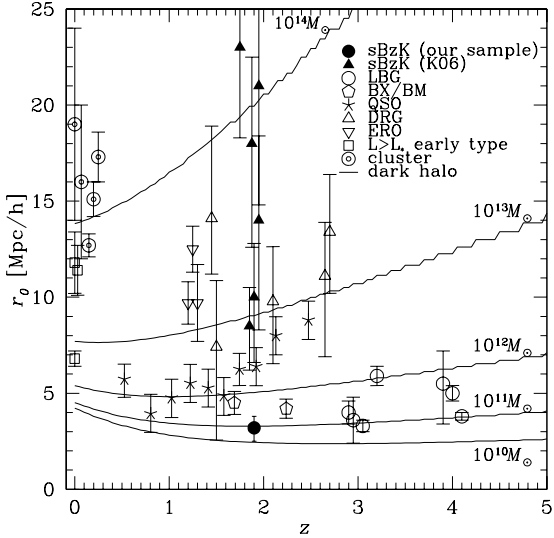}
  \caption{Correlation lengths of various types of objects 
   over $0 \le z \le 4$.
   The filled symbols represent sBzKs; circle: our sample, 
   triangles: K06.
   The other symbols indicate the measurements for 
   other objects taken from the literature; 
   open circles: LBGs at $z \sim 3$ ($23.5<R<25.5$, \citet{ade05};\  
   $20.0<I<24.5$, \citet{fou03}) and $z \sim 4$ ($i'<25,26,27.5$, \citet{ouc05}), 
   open pentagons: BX/BM galaxies with $23.5<R<25.5$ at $z \sim 2$ \citep{ade05}, 
   stars: QSOs at $z \sim 0.5$ --- 2 
   ($18.25<b_J<20.85$, \citet{cro05}), 
   open triangles: Distant Red Galaxies (DRGs) 
   at $z \sim 2$ --- 3 ($\KVega<21$, \citet{qua06};\  
   $K<20.7$, \citet{fou06};\ $K<23.5$, \citet{gra06}), 
   open inverted triangles: Extremely Red Objects (EROs) 
   at $z \sim 1$ --- 2 ($\KVega<24$, \citet{dad03};\
   $\KVega<22$, \citet{roc03};\ $\KVega<18.4$, \citet{bro05}), 
   open squares: present-day luminous early-type galaxies \citep{ove03}, 
   double circles: present-day clusters of galaxies \citep{bah03}. 
   The five curves show the correlation length of dark haloes 
   with five different minimum masses as labeled. 
   }
  \label{r0_z}
  \end{center}
\end{figure}

Figure \ref{r0_z} compares the correlation lengths of sBzKs 
with those of various types of objects from  
present-day early-type galaxies and clusters of galaxies, up to
galaxies at $z = 4$.
The five curves show the correlation lengths of dark haloes 
with five fixed $\MDHmin$ values as a function of redshift. 
Haloes with a fixed minimum mass have larger $r_0$ at higher $z$ 
because $b$ increases with $z$ for any $\MDH$.

It is found from Figure \ref{r0_z} that at $z \sim 2$ 
faint ($K<23.2$) sBzKs and BX/BM galaxies 
have similar correlation lengths and thus similar dark-halo 
masses of the order of $\sim 10^{11} M_\odot$. 
This result is in accord with the recent claim by \citet{red05} 
that a high fraction ($70$ -- $80\%$) of BXs/BMs 
with $22.3 < K < 22.8$ are also selected as sBzKs 
and that the fraction increases with decreasing $K$ brightness.
Therefore, faint sBzKs and BXs/BMs are largely overlapping 
subsets of star-forming galaxies at $z \sim 2$ 
hosted by relatively low-mass haloes.
In Figure \ref{r0_z}, LBGs at $z = 3$ -- 4 have 
also similar masses to faint sBzKs, 
while haloes hosting QSOs are on average 
an order of magnitude more massive.

Brighter sBzKs with $K\simeq 21$ -- $22$ are found to 
have as large correlation lengths, 
$r_0 \simeq 10$ -- 15 $h^{-1}$ Mpc, 
as DRGs at $z \sim 2$ -- 3 and EROs at $z \sim 1$ -- 2.
They are located in haloes as massive as 
the order of $\MDH \sim 10^{13} M_\odot$.
Their correlation lengths partly overlap with those of 
present-day luminous early-type galaxies and clusters.
Large fractions of DRGs and EROs are known to be passive galaxies, 
while sBzKs are star-forming galaxies.
This implies that haloes with similar masses can host galaxies 
with extremely different star-formation properties.
Brightest sBzKs with $K<21$ are most strongly clustered 
among the high-$z$ populations plotted here. 
Their correlation lengths are comparable to or even larger 
than those of present-day rich clusters.

\subsection{Present-day Descendants of sBzKs}

We use the extended Press-Schechter formalism 
\citep{bon91,bow91}
to predict the mass of the present-day descendants of 
dark haloes hosting sBzKs at $z \sim 2$.
We find that haloes at $z=2$ with a mass equal to 
the $\MDHav$ of faint ($K<23.2$) sBzKs 
($2.8 \times 10^{11} M_\odot$) 
will become haloes with masses of 
$(3.7-10.0) \times 10^{11} M_\odot$ at $z=0$ 
($68\%$ range of the distribution function; 
the extended Press-Schechter formalism 
can predict descendant masses only in a statistical sense).\footnotemark[3]
\footnotetext[3]{Dr. Takashi Hamana kindly provided the code to calculate 
the distribution function.}
Thus, the present-day descendants of the typical haloes 
of faint sBzKs are likely to be less massive than that of the Milky Way.
They may host sub-$L^*$ galaxies in nearby fields.
On the basis of a clustering analysis of BX/BM galaxies, 
\citet{ade05} concluded that BXs/BMs are progenitors of normal 
ellipticals in the local universe. 
The correlation length they obtained is 
$r_0 = 4.2 \pm 0.5$ $h^{-1}$Mpc for BXs 
and $r_0 = 4.5 \pm 0.6$ $h^{-1}$Mpc for BMs. 
These values correspond to dark haloes several times 
more massive than those of our faint sBzKs. 
Thus, their conclusion does not seem to conflict with our result 
obtained here.

On the other hand, haloes with a mass equal to 
the $\MDHav$ of $K < 20.4$ sBzKs in K06 
($1.6 \times 10^{14} M_\odot$) are predicted to have 
$(4.3-9.3) \times 10^{14} M_\odot$ at present.
These masses are comparable to those of most massive clusters 
like Coma.
The number density of sBzKs with $K < 20.4$, 
$(1-2) \times 10^{-5} \rm{Mpc^{-3}}$, is close to 
that of the present-day rich clusters 
more massive than $2 \times 10^{14} M_{\sun}$, 
$\sim 1 \times 10^{-5} \rm{Mpc^{-3}}$ \citep{rin06}. 
This suggests that brightest sBzKs are ancestors of 
central galaxies in rich clusters seen at present.

These calculations suggest that sBzKs evolve into 
galaxies over a wide range of mass (or luminosity) 
in a variety of environment at $z=0$, depending on 
their apparent $K$ brightness.

\section{Conclusions}
\label{sec;conclusion}
In this paper, we have studied clustering properties of 
star-forming BzK galaxies (sBzKs) at $1.4 \lesssim z \lesssim 2.5$
over a wide range of $K$ brightness ($K<23.2$).

We have used deep multi-color data of 180 arcmin$^2$ 
in the Subaru Deep Field to construct a sample of 
1092 faint ($K<23.2$) sBzK galaxies.
We have derived the angular correlation function (ACF) 
of the sBzKs, and measured its amplitude to be 
$A = (0.58 \pm 0.13) $ [arcsec$^{0.8}$] by fitting a power law of 
$w(\theta) \propto \theta^{-0.8}$ to the data.
We have then transformed the ACF amplitude into the correlation 
length assuming a Gaussian redshift distribution of 
$z_c=1.9$ and $\sigma_z = 0.35 \pm 0.1$, 
and obtained $r_0 = 3.2^{+0.6}_{-0.7}$ $h^{-1}$ Mpc. 
We have not been able to derive the ACF of passive BzKs in our data 
because of the small sample size.
We infer from the correlation length that our sBzKs reside in 
haloes with an average mass of $2.8 \times 10^{11} M_\odot$ 
and a minimum mass of $1.0 \times 10^{11} M_\odot$.

We then have inferred dark-halo masses for six bright ($K<21.9$) 
sBzK samples of K06 in the same manner.
Combining our data with those of K06, 
we have examined how the mass of hosting dark haloes 
depends on the $K$ luminosity of sBzKs.
We have found that the mass of dark haloes rapidly increases
with $K$ brightness of individual sBzKs; 
K06's sBzKs are brighter than ours 
by only up to $\simeq 10$ times, 
but they are hosted by haloes two to three orders of magnitude 
more massive than the haloes of our sBzKs.
The halo occupation number, the number of BzKs hosted in 
a dark halo, is found to be higher for brighter sBzKs.

The correlation lengths of sBzKs have been 
compared with those of various types of objects up to $z=4$.
We have found that faint ($K<23.2$) sBzKs have similar $r_0$ values 
to BX/BM galaxies, which are optically-selected star-forming 
galaxies at $z \sim 2$. 
We argue that these two types of galaxies are similar populations 
hosted by relatively low-mass haloes 
with the order of $\MDH \sim 10^{11} M_\odot$.
On the other hand, sBzKs with $K \simeq 21-22$, EROs, and DRGs
are found to reside in massive haloes 
with the order of $\sim 10^{13} M_\odot$.
The clustering of brightest sBzKs with $K<21$ is the strongest 
among the high-$z$ populations.

Finally, we have predicted present-day descendants of haloes 
hosting sBzKs using the extended Press-Schechter formalism.
Descendants are found to span a wide range of mass, 
depending on the $K$ brightness of sBzKs in them.
Typical descendants of haloes hosting sBzKs with $K<23.2$ 
are less massive than the Milky Way, 
while those for $K<20.4$ sBzKs are as massive as richest clusters.

\acknowledgments
We wish to thank Takashi Hamana for providing his code 
to compute the mass growth of dark haloes using
the extended Press-Schechter formalism.
We are very grateful to Emanuele Daddi for kindly providing 
new data of the redshift distribution of 81 sBzKs prior to publication.
We also thank an anonymous referee for useful 
comments which have greatly improved the paper.
The United Kingdom Infrared Telescope is operated by the Joint Astronomy 
Centre on behalf of the U.K. Particle Physics and Astronomy Council.
The WFCAM data were reduced on the general common-use 
computer system at the Astronomy Data Center, ADC, 
of the National Astronomical Observatory of Japan.

\end{document}